# Analytical construction of the projectile motion trajectory in midair

**Peter Chudinov, Vladimir Eltyshev and Yuri Barykin**

Department of Engineering, Perm State Agro-Technological University, 614990, Perm, Russia

E-mail: chupet@mail.ru

**Abstract.** A classic problem of the motion of a projectile thrown at an angle to the horizon is studied. Air resistance force is taken into account. The quadratic law for the resistance force is used. An analytic approach applies for the investigation. Equations of the projectile motion are solved analytically for an arbitrarily large period of time. The constructed analytical solutions are universal. As a limit case of motion, the vertical asymptote formula is obtained. The value of the vertical asymptote is calculated directly from the initial conditions of motion. There is no need to study the problem numerically. The found analytical solutions are highly accurate over a wide range of parameters. The motion of a baseball, a tennis ball and shuttlecock of badminton are presented as examples.

## 1. Introduction

The problem of the motion of a projectile in midair is of great interest for investigators for centuries. There are a lot of publications on this problem. It is almost impossible to make a detailed review of all published articles on this topic. Together with numerical methods, attempts are still being made to obtain the analytical solutions. Many such solutions of a particular type have been obtained. They are valid for limited values of the physical parameters of the problem (for the linear law of the medium resistance at low speeds, for short travel times, for low, high and split angle trajectory regimes). Both the traditional approaches, and the modern methods are used for the construction of the analytical solutions. But these proposed approximate analytical solutions are rather complicated and inconvenient for educational purposes. Some approximate solutions use special functions, for example, the Lambert W function. So the description of the projectile motion by means of simple approximate analytical formulas under the quadratic air resistance has great methodological and educational importance.

The purpose of the present original work is to give simple formulas for the construction of the trajectory of the projectile motion with quadratic air resistance for an arbitrarily large period of time. It is well known that the projectile trajectory has a vertical asymptote in the resistant medium. Construction of the trajectory of the projectile over an arbitrarily large period of time allows us to determine analytically one of the important characteristics of the motion - the value of the vertical asymptote. The subject of research in the proposed paper has something in common with the content of [1,2]. In these papers, analytical formulas were obtained for the value of the vertical asymptote of the projectile trajectory. However, a significant difference between the results of this paper and the above-mentioned papers is that the vertical asymptote formula obtained in this research has a much higher accuracy. In this article, the construction of the trajectory is carried out using the approach [3]. This approach allows to construct a trajectory of the projectile with the help of elementary functions without using numerical schemes. Following other authors, we call this approach the analytic approach. From the point of view of the applied methods of solution, the present research is a development of the approaches used in [4,5]. The conditions of applicability of the quadratic resistance law are deemed to be fulfilled, i.e. Reynolds number $Re$ lies within $1\times10^3 < Re < 2\times10^5$.

## 2. Equations of projectile motion

Here we state the formulation of the problem and the equations of the motion [5]. Suppose that the force of gravity affects the projectile together with the force of air resistance $\mathbf{R}$ (see figure 1). Air resistance force is proportional to the square of the velocity $V$ of the projectile and is directed opposite the velocity vector. For the convenience of further calculations, the drag force will be written as $R = mgkV^2$. Here $m$ is the mass of the projectile, $g$ is the acceleration due to gravity, $k$ is the proportionality factor. In most papers, the motion of the projectile is studied in projections on the Cartesian axis. Meanwhile, the equations of motion of the projectile in projections on the natural axes, often used in ballistics [6], are very useful. They have the following form

$$\frac{dV}{dt} = -g\sin\theta - gkV^2, \quad \frac{d\theta}{dt} = -\frac{g\cos\theta}{V}, \quad \frac{dx}{dt} = V\cos\theta, \quad \frac{dy}{dt} = V\sin\theta. \qquad (1)$$

Here $V$ is the velocity of the projectile, $\theta$ is the angle between the tangent to the trajectory of the projectile and the horizontal, $x, y$ are the Cartesian coordinates of the projectile,

$$k = \frac{\rho_a c_d S}{2mg} = \frac{1}{V_{term}^2} = const,$$

$\rho_a$ is the air density, $c_d$ is the drag factor for a sphere, $S$ is the cross-section area of the object, and $V_{term}$ is the terminal velocity. The first two equations of the system (1) represent the projections of the vector equation of motion on the tangent and principal normal to the trajectory, the other two are kinematic relations connecting the projections of the velocity vector projectile on the axis $x, y$ with derivatives of the coordinates.

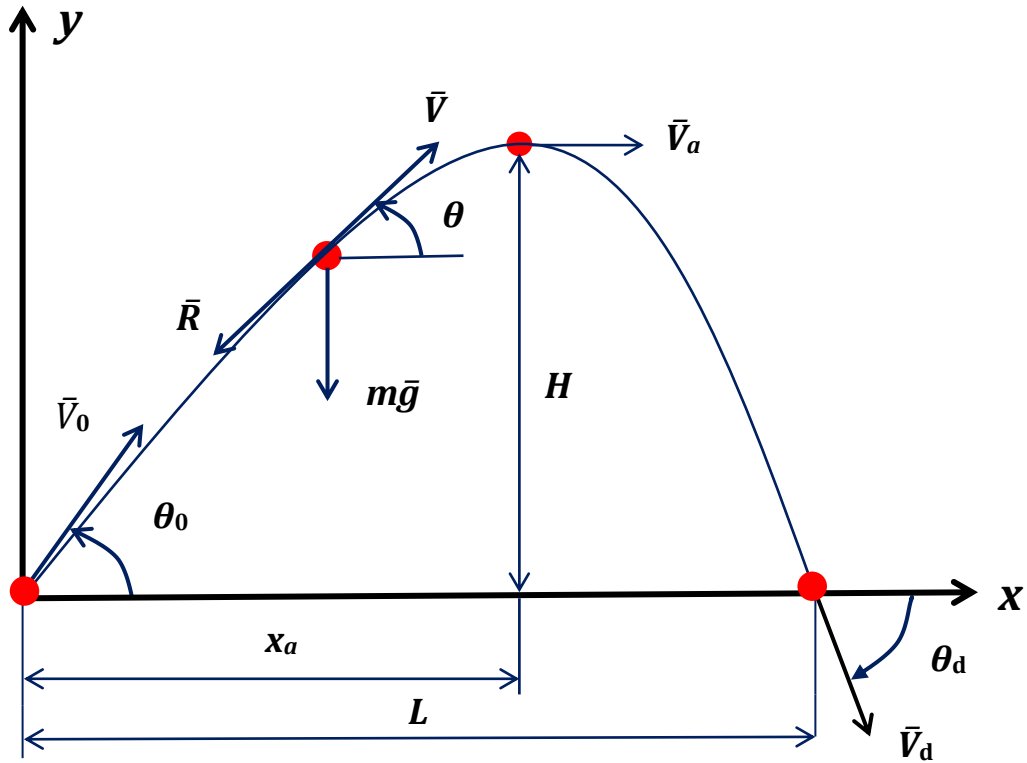

**Figure 1.** Basic motion parameters.

The well-known solution of system (1) consists of an explicit analytical dependence of the velocity on the slope angle of the trajectory and three quadratures

$$V(\theta) = \frac{V_0 \cos\theta_0}{\cos\theta \sqrt{1 + kV_0^2 \cos^2\theta_0 (f(\theta_0) - f(\theta))}}, \quad f(\theta) = \frac{\sin\theta}{\cos^2\theta} + \ln\tan\left(\frac{\theta}{2} + \frac{\pi}{4}\right), \quad (2)$$

$$x = x_0 - \frac{1}{g}\int_{\theta_0}^{\theta} V^2 d\theta, \quad y = y_0 - \frac{1}{g}\int_{\theta_0}^{\theta} V^2 \tan\theta\, d\theta, \quad t = t_0 - \frac{1}{g}\int_{\theta_0}^{\theta} \frac{V}{\cos\theta} d\theta. \quad (3)$$

Here $V_0$ and $\theta_0$ are the initial values of the velocity and of the slope of the trajectory respectively, $t_0$ is the initial value of the time, $x_0$, $y_0$ are the initial values of the coordinates of the projectile (usually accepted $t_0 = x_0 = y_0 = 0$). The derivation of the formulae (2) is shown in the well-known monograph [7]. The integrals on the right-hand sides of formulas (3) cannot be expressed in terms of elementary functions. Hence, to determine the variables $t$, $x$ and $y$ we must either integrate system (1) numerically or evaluate the definite integrals (3). In [5], the integrals were calculated in elementary functions over most of the angle $\theta$.

The purpose of this study is to calculate the integrals (3) in elementary functions with the necessary accuracy over the entire interval of variation of the variable $\theta$: [ $-\pi/2 \leq \theta \leq \theta_0$ ].

## 3. The obtaining an analytical solutions of the problem

To understand this article, we state the following moments. In [3], a very fruitful idea was proposed for calculating integrals (3). Based on this idea, in [4,5] there were obtained high-precision analytical solutions for the variables $x, y$ and $t$ over a sufficiently large interval of variation of the variable $\theta$. In the present work, these solutions are continued up to the limit value $\theta = -\pi/2$.

The idea [3] is as follows. The task analysis shows, that equations (3) are not exactly integrable owing to the complicated nature of function (2)

$$f(\theta) = \frac{\sin\theta}{\cos^2\theta} + \ln\tan\left(\frac{\theta}{2} + \frac{\pi}{4}\right).$$

The odd function $f(\theta)$ is defined in the interval $-\pi/2 < \theta < \pi/2$. Therefore, it can be assumed that a successful approximation of this function will make it possible to calculate analytically the definite integrals (3) with the required accuracy. In [3] presents a simple approximation in the mathematical sense of a function $f(\theta)$ by a second-order polynomial of the following form (polynomial is with respect to a function $\tan\theta$)

$$f_a(\theta) = a_1 \tan\theta + b_1 \tan^2\theta.$$

The function $f_a(\theta)$ well approximates the function $f(\theta)$ only on the specified interval $[0, \theta_0]$, since the function $f_a(\theta)$ contains an even term. Under the condition $\theta < 0$, another approximation is required because the function $f(\theta)$ is odd.

In [4,5] two approximations of the function $f(\theta)$ on the whole interval $-\pi/2 < \theta < \pi/2$ were proposed. The first approximation uses a second-order polynomial, the second approximation uses a third-order polynomial. In this paper, the first approximation is used, as a simpler one. Approximation of the function $f(\theta)$ by a second order polynomial $f_2(\theta)$ has the following form

$$f_2(\theta) = \begin{cases} \alpha_1 \tan\theta + \alpha_2 \tan^2\theta, & \text{on condition} \quad \theta \geq 0, \\ \alpha_1 \tan\theta - \alpha_2 \tan^2\theta, & \text{on condition} \quad \theta \leq 0. \end{cases} \qquad (4)$$

Approximation (4) is two-parameter. The parameters are the coefficients $\alpha_1$ and $\alpha_2$. This approximation also satisfies the condition $f_2(0) = f(0) = 0$. The coefficients $\alpha_1$ and $\alpha_2$ can be chosen in such a way as to smoothly connect the functions $f(\theta)$ and $f_2(\theta)$ to each other with the help of conditions

$$f_2(\theta_0) = f(\theta_0), \quad f_2'(\theta_0) = f'(\theta_0). \qquad (5)$$

From conditions (5) we have

$$\alpha_1 = \frac{2\ln\tan(\theta_0/2 + \pi/4)}{\tan\theta_0}, \quad \alpha_2 = \frac{1}{\sin\theta_0} - \frac{\ln\tan(\theta_0/2 + \pi/4)}{\tan^2\theta_0}.$$

Note that due to the oddness of the functions $f(\theta)$ and $f_2(\theta)$ equality is held $f_2(-\theta_0) = f(-\theta_0)$. This function $f_2(\theta)$ well approximate the function $f(\theta)$ over a sufficiently large interval of its definition for any values $\theta_0$ (see figure 3).

Based on approximation (4), in [5] analytical formulas for the variables $x, y$ и $t$ are obtained by calculating integrals (3). The results [5] are substantially used in this paper. It is quite necessary to quote the formulas [5] in this article, because without them it impossible to use the results obtained in this paper. Therefore, we reproduce the needed formulas from [5] :

$$x_1(\theta) = -\frac{1}{g}\int_{\theta_0}^{\theta} V^2 d\theta = A_1 \arctan\left(\frac{2b_2 \tan\theta + 1}{b_3}\right)\Bigg|_{\theta_0}^{\theta} \quad \text{in case of} \quad \theta \geq 0,$$

$$x_2(\theta) = -A_2 \arctan\left(\frac{2b_2 \tan\theta - 1}{b_4}\right)\Bigg|_{\theta_0}^{\theta} \quad \text{in case of} \quad \theta \leq 0.$$

The following notations are introduced here:

$$A_1 = \frac{2}{gk\alpha_1 b_3}, \quad A_2 = \frac{2}{gk\alpha_1 b_4}, \quad b_1 = \left(\frac{1}{kV_0^2 \cos^2\theta_0} + f(\theta_0)\right)/\alpha_1, \quad b_2 = \frac{\alpha_2}{\alpha_1},$$

$$b_3 = \sqrt{-1 - 4b_1 b_2}, \quad b_4 = \sqrt{-1 + 4b_1 b_2}.$$

Thus, the dependence $x(\theta)$ has the following form:

$$x(\theta) = x_1(\theta) - x_1(\theta_0) \qquad \text{in case of} \quad \theta \geq 0,$$
$$x(\theta) = x_1(0) - x_1(\theta_0) + x_2(\theta) - x_2(0) \qquad \text{in case of} \quad \theta \leq 0. \qquad (6)$$

For the coordinate $y$ we obtain:

$$y_1(\theta) = -\frac{1}{g}\int_{\theta_0}^{\theta} V^2 \tan\theta d\theta = \left(-B_1 \arctan\left(\frac{1 + 2b_2 \tan\theta}{b_3}\right) + B_2 \ln\left|-b_1 + \tan\theta + b_2 \tan^2\theta\right|\right)\Bigg|_{\theta_0}^{\theta}$$

in case of $\theta \geq 0$,

$$y_2(\theta) = \left. \left( -B_3 \arctan\left(\frac{-1+2b_2 \tan\theta}{b_4}\right) - B_2 \ln\left|b_1 - \tan\theta + b_2 \tan^2\theta\right| \right) \right|_{\theta_0}^{\theta} \quad \text{in case of } \theta \leq 0.$$

The following notations are introduced here:

$$B_1 = 1/(kg\alpha_2 b_3), \quad B_2 = 1/(2kg\alpha_2), \quad B_3 = 1/(kg\alpha_2 b_4).$$

Thus, the dependence $y(\theta)$ has the following form:

$$y(\theta) = y_1(\theta) - y_1(\theta_0) \quad \text{in case of } \theta \geq 0,$$
$$y(\theta) = y_1(0) - y_1(\theta_0) + y_2(\theta) - y_2(0) \quad \text{in case of } \theta \leq 0. \quad (7)$$

The variable $t$ is defined by the formulas

$$t_1(\theta) = \left. \frac{1}{g\sqrt{k\alpha_2}} \arctan\left[\frac{(1+2b_2\tan\theta)\sqrt{b_1 - \tan\theta - b_2\tan^2\theta}}{2\sqrt{b_2}\left(-b_1 + \tan\theta + b_2\tan^2\theta\right)}\right] \right|_{\theta_0}^{\theta} \quad \text{in case of } \theta \geq 0,$$

$$t_2(\theta) = \left. -\frac{1}{g\sqrt{k\alpha_2}} \ln\left|-1 + 2b_2\tan\theta + 2\sqrt{b_2}\sqrt{b_1 - \tan\theta + b_2\tan^2\theta}\right| \right|_{\theta_0}^{\theta} \quad \text{in case of } \theta \leq 0.$$

Thus, the dependence $t(\theta)$ has the following form:

$$t(\theta) = t_1(\theta) - t_1(\theta_0) \quad \text{in case of } \theta \geq 0,$$
$$t(\theta) = t_1(0) - t_1(\theta_0) + t_2(\theta) - t_2(0) \quad \text{in case of } \theta \leq 0. \quad (8)$$

Consequently, the basic functional dependencies of the problem $x(\theta), y(\theta), t(\theta)$ are written in terms of elementary functions. We note that formulas (6) – (8) also define the dependences $y = y(x)$, $y = y(t)$, $x = x(t)$ in a parametric way. The obtained formulas very well describe the trajectory of the projectile in the range of variation of the angle $\theta$, corresponding to the condition $y \geq 0$. An example of the use of formulas (6) – (8) is shown in figure 2.

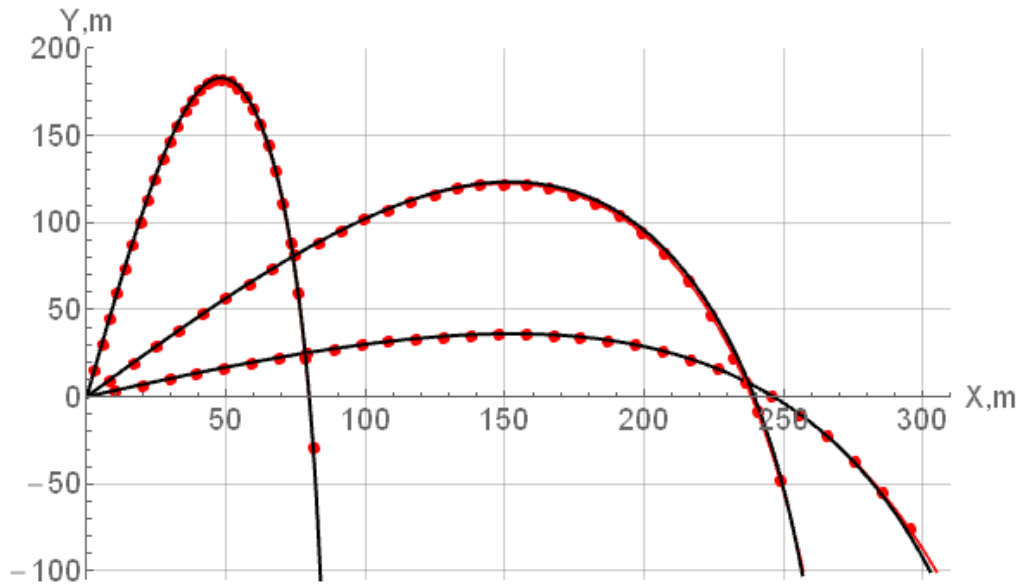

**Figure 2.** The graphs of the trajectory $y = y(x)$ at launching angles $\theta_0 = 20°, 50°, 80°$, $V_0 = 80$ m/s.

Figure 2 presents the results of plotting the projectile trajectories with the aid of formulas (6) – (8) over a wide range of the change of the initial angle $\theta_0$. The following parameters values are used

$$k = 0.000625 \ \text{s}^2/\text{m}^2, \quad g = 9.81 \ \text{m/s}^2.$$

The used value of the parameter $k$ is the typical value of the baseball drag coefficient [1]. The thick solid lines in figure 2 are obtained by numerical integration of system (1) with the aid of the 4-th order Runge-Kutta method. The red dots lines are obtained with using analytical formulas (6) – (8). As it can be seen from figure 2, formulas (6) – (8) with high accuracy approximate the trajectory of the projectile in a fairly wide range of angle $\theta$, satisfying the condition $y \geq 0$.

## 4. Analytical solutions for the approximation function $f(\theta)$ as the trajectory approaches the asymptote

Thus, analytical solutions are constructed in [5] for the projectile coordinates $x, y$ for any values of the angle of inclination of the trajectory $\theta$. However, at $\theta \to -\pi/2$, the numerical and analytical solutions begin to diverge. This can be seen even in figure 2. In order to prevent this, we proceed as follows. We divide the full interval of the change in the angle $\theta$ of the slope of the trajectory ($-\pi/2 \leq \theta \leq \theta_0$] into two intervals:

$$(-\pi/2 \leq \theta \leq \theta_1] \ \text{and} \ [\theta_1 \leq \theta \leq \theta_0].$$

We choose the value $\theta_1$ taking the following considerations into account. Since the odd function $f_2(\theta)$ approximates well the trajectory of the projectile in the segment $[-\theta_0 \leq \theta \leq \theta_0]$ and even further, we take the middle of the gap $(-\pi/2 \leq \theta \leq -\theta_0]$ as the value $\theta_1$. In that way $\theta_1 = -\theta_0/2 - \pi/4$. Now we will approximate the function $f(\theta)$ on the interval ($-\pi/2 \leq \theta \leq \theta_1$] by the function

$$f_3(\theta) = \beta_0 + \beta_1 \tan\theta - \beta_2 \tan^2\theta.$$

In contrast to the two-parameter approximation (4), this approximation is three-parameter. The presence of the third parameter $\beta_0$ allows us to impose an additional condition on the function $f_3(\theta)$ in addition to the conditions (5). We impose following conditions on the function $f_3(\theta)$

$$f_3(\theta_1) = f(\theta_1), \quad f_3'(\theta_1) = f'(\theta_1), \quad f_3(-89°) = f(-89°). \tag{9}$$

The first two conditions are similar to conditions (5), the third condition ensures that the trajectory approaches the asymptote. From conditions (9) we find the coefficients $\beta_0, \beta_1, \beta_2$:

$$\beta_2 = \frac{\left(f(\theta_1) + f(89°)\right)\cos\theta_1 - 2\left(\tan\theta_1 + \tan 89°\right)}{(\tan\theta_1 + \tan 89°)^2 \cos\theta_1}, \quad \beta_1 = \frac{2(1 + \beta_2 \sin\theta_1)}{\cos\theta_1},$$

$$\beta_0 = f(\theta_1) - \beta_1 \tan\theta_1 + \beta_2 \tan^2\theta_1.$$

Figure 3 shows the results of approximating the function $f(\theta)$ by the functions $f_2(\theta)$ and $f_3(\theta)$. The solid green line is a graph of the function $f(\theta)$. The red dotted line is a graph of the function $f_2(\theta)$ in the gap $[\theta_1 \leq \theta \leq \theta_0]$. The black dotted line is a graph of the function $f_3(\theta)$ in the gap ($-\pi/2 \leq \theta \leq \theta_1$]. As it can be seen from figure 3, the used functions $f_2(\theta)$ and $f_3(\theta)$ approximate well the function $f(\theta)$ over the entire interval ($-\pi/2 \leq \theta \leq \theta_0$]. When plotting the functions $f(\theta), f_2(\theta), f_3(\theta)$ in figure 3, the initial value is used $\theta_0 = 60°$.

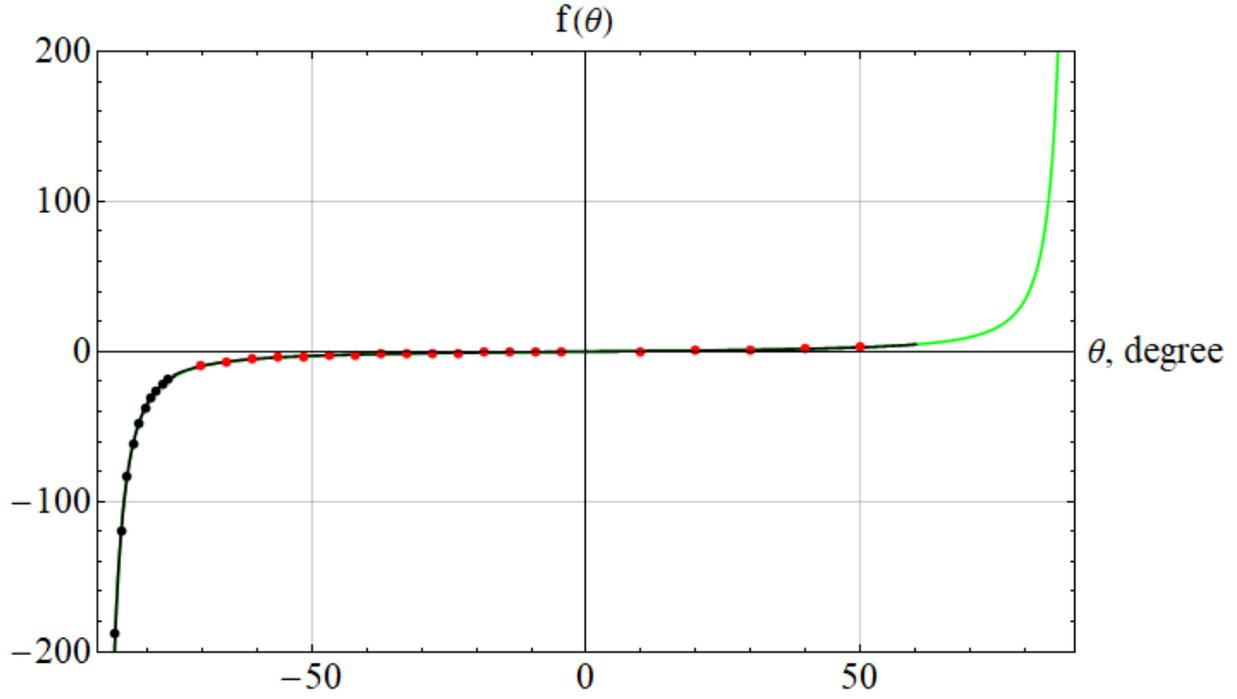

**Figure 3.** Approximation of the function $f(\theta)$ by functions $f_2(\theta)$ and $f_3(\theta)$.

Now we integrate the first of the integrals (3). For the coordinate $x$ we obtain:

$$x_3(\theta) = x_0 - A_3 \arctan\left(\frac{2d_1 \tan\theta - 1}{d_2}\right)\Bigg|_{\theta_1}^{\theta},$$

где $d_0 = \dfrac{1}{\beta_1}\left(\dfrac{1}{kV_0^2 \cos^2\theta_0} + f(\theta_0) - \beta_0\right)$, $d_1 = \dfrac{\beta_2}{\beta_1}$, $d_2 = \sqrt{4d_0 d_1 - 1}$, $A_3 = \dfrac{2}{gk\beta_1 d_2}$.

Integrating the second integral (3), we obtain the dependence $y(\theta)$:

$$y_3(\theta) = y_0 - \left\{\frac{1}{gk\beta_2 d_2}\arctan\left(\frac{2d_1 \tan\theta - 1}{d_2}\right) - \frac{1}{2gk\beta_2}\ln\left(d_0 - \tan\theta + d_1 \tan^2\theta\right)\right\}\Bigg|_{\theta_1}^{\theta}.$$

The third integral (3) gives the dependence $t(\theta)$:

$$t_3(\theta) = t_0 - \frac{1}{g\sqrt{k\beta_2}}\ln\left|1 - 2d_1\tan\theta - 2\sqrt{d_1}\sqrt{d_0 - \tan\theta + d_1 \tan^2\theta}\right|\Bigg|_{\theta_1}^{\theta}.$$

The obtained formulas $x_3(\theta), y_3(\theta)$ for the $x, y$ coordinates allow to construct analytically the projectile trajectory over the entire interval of variation of the angle of slope $\theta$.

## 5. Construction of the trajectory. Formula for the vertical asymptote. The results of the calculations

Now we can analytically construct the trajectory of the projectile for an arbitrarily long period of time. However, instead of an unlimited interval $[0 \leq t < \infty)$ for time $t$, it is more convenient to use a limited interval ($-\pi/2 \leq \theta \leq \theta_0$] for the angle $\theta$.

Thus, the projectile trajectory over the entire interval of the variation of the angle $\theta$ $(-\pi/2 \leq \theta \leq \theta_0]$ is described by the equations:

on the interval $[0 \leq \theta \leq \theta_0]$

$$t(\theta) = t_1(\theta) - t_1(\theta_0),$$
$$x(\theta) = x_1(\theta) - x_1(\theta_0),$$
$$y(\theta) = y_1(\theta) - y_1(\theta_0);$$

on the interval $[\theta_1 \leq \theta \leq 0]$

$$t(\theta) = t_1(0) - t_1(\theta_0) + t_2(\theta) - t_2(0),$$
$$x(\theta) = x_1(0) - x_1(\theta_0) + x_2(\theta) - x_2(0),$$
$$y(\theta) = y_1(0) - y_1(\theta_0) + y_2(\theta) - y_2(0);$$

on the interval $(-\pi/2 \leq \theta \leq \theta_1]$

$$t(\theta) = t_1(0) - t_1(\theta_0) + t_2(\theta_1) - t_2(0) + t_3(\theta) - t_3(\theta_1),$$
$$x(\theta) = x_1(0) - x_1(\theta_0) + x_2(\theta_1) - x_2(0) + x_3(\theta) - x_3(\theta_1),$$
$$y(\theta) = y_1(0) - y_1(\theta_0) + y_2(\theta_1) - y_2(0) + y_3(\theta) - y_3(\theta_1). \qquad (10)$$

In contrast to all previously obtained solutions to the problem of the movement of a projectile in a medium with a quadratic law of resistance, formulas (10) allow us to construct the trajectory of the projectile for any initial conditions of throwing $V_0$, $\theta_0$ and for any values of the drag coefficient $k$.

An example of dependencies $x(\theta), y(x)$ constructed using equations (10) is shown in figure 4. In the calculations, the following parameter values were used:

$$V_0 = 120 \text{ m/s}, \quad k = 0.000625 \text{ s}^2/\text{m}^2, \quad g = 9.81 \text{ m/s}^2, \quad \theta_0 = 45°.$$

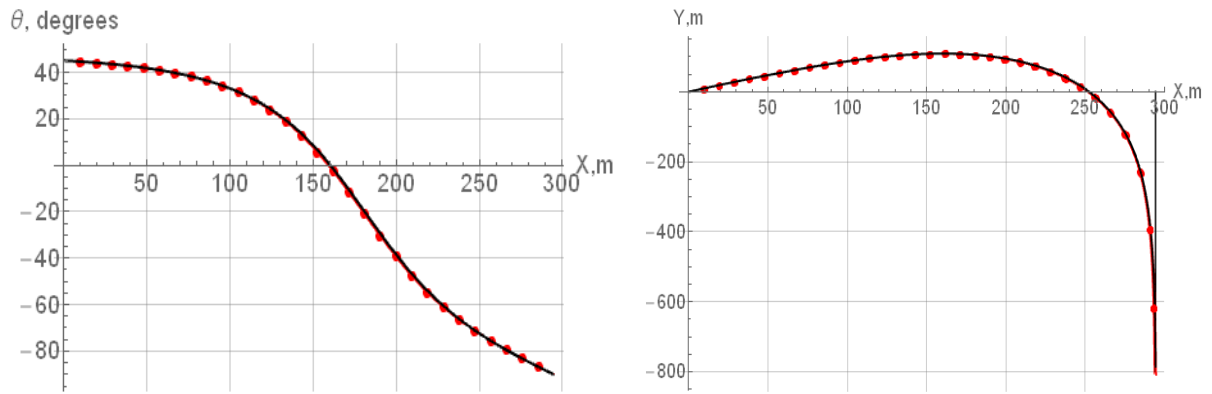

**Figure 4.** The graphs of the functions $x(\theta), y(x)$.

The thick solid black lines in figure 4 are obtained by numerical integration of system (1) with the aid of the 4-th order Runge-Kutta method. The red dots lines are obtained with using analytical formulas (10). As it can be seen from figure 4, the analytical solutions (dotted lines) and numerical solutions are the same over the entire interval of change of the angle of inclination of the trajectory $(-\pi/2 \leq \theta \leq \theta_0]$. Each of the graphs allows us to approximately determine the value of the asymptote $x_{as}$ of the projectile trajectory. The exact value of the asymptote is determined by the expression

$$x_{as} = x_1(0) - x_1(\theta_0) + x_2(\theta_1) - x_2(0) + x_3(-\pi/2) - x_3(\theta_1).$$

Using the previously obtained formulas, we write the final analytical expression for the asymptote of the projectile trajectory

$$x_{as} = A_1 \arctan\left(\frac{b_3}{2b_1 \cot\theta_0 - 1}\right) + A_2 \arctan\left(\frac{b_4}{1 - 2b_1 \cot\theta_1}\right) + A_3 \operatorname{arccot}\left(\frac{1 - 2d_1 \tan\theta_1}{d_2}\right). \quad (11)$$

Multipliers $A_1, A_2, A_3$ and coefficients $b_1, b_3, b_4, d_1, d_2$ were introduced earlier. Note that the asymptote value is calculated directly from the initial conditions of motion $V_0, \theta_0$, without integrating the equations of motion of the projectile. In figure 3, the value of the asymptote determined by the numerical integration of the system of equations of motion of the projectile (1) is $x_{as} = 294.7$ m. The asymptote is shown by a solid vertical black line. Formula (11) gives a value of $x_{as} = 293.9$ m. The error in calculating the asymptote is 0.28%.

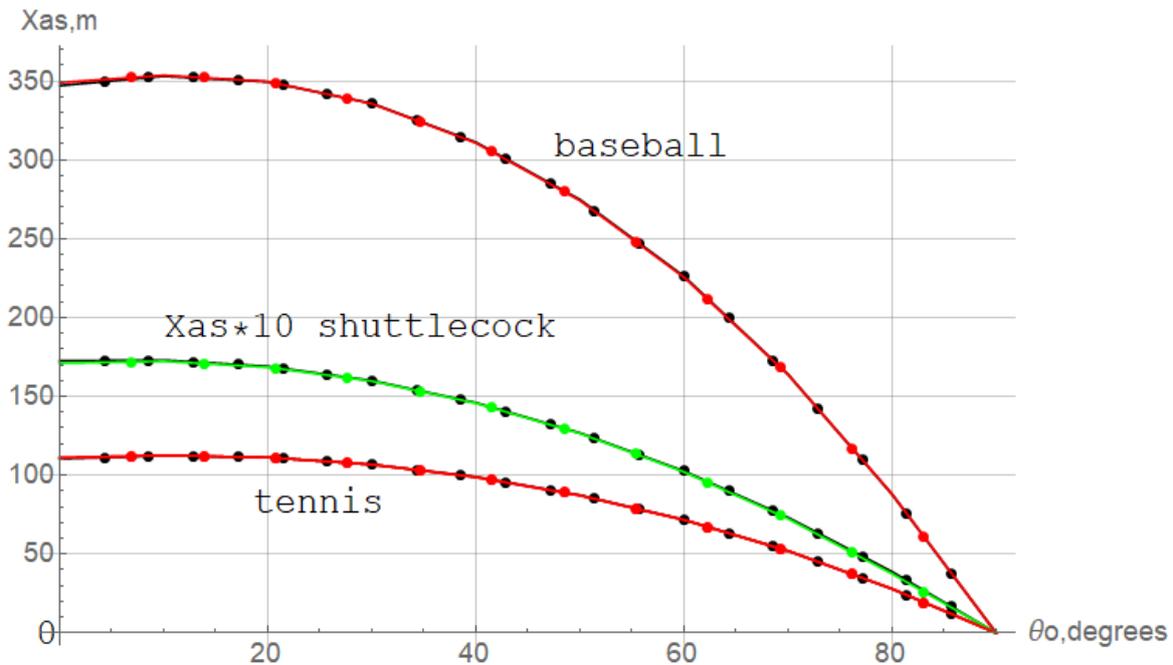

**Figure 5.** The graph of the function $x_{as}(\theta_0)$.

Figure 5 shows the dependence $x_{as} = x_{as}(\theta_0)$. The upper curve in figure 5 is plotted for parameter values $k = 0.000625$ s$^2$/m$^2$, $V_0 = 120$ m/s (value of $k$ for a baseball), bottom curve is plotted at values $k = 0.002$ s$^2$/m$^2$, $V_0 = 70$ m/s ($k$ value for tennis ball). The middle curve is plotted for shuttlecock of badminton ($k = 0.022$ s$^2$/m$^2$, $V_0 = 100$ m/s). In this case, the values of $x_{as}$ are increased 10 times. The black dots lines in figure 5 are obtained by numerical integration of system (1) with the aid of the 4-th order Runge-Kutta method. The red and green dots lines are obtained using analytical formula (11). Black and red (green) curves completely coincide over the entire interval of variation of the initial projectile throw angle $0° \leq \theta_0 \leq 90°$. It should be noted that the used values of the resistance coefficient $k$ in figure 5 vary 35 times. This indicates a high accuracy of formula (11) in a wide range of variation of the parameter $k$.

## 6. Comparison of results

Note that an approximate formulas for the vertical asymptote of the projectile were proposed in [1,2]. The vertical asymptote formula in [1] has the following form (in the notations of this article)

$$x_{as} = \frac{\cos\theta_0}{gk}\left(\frac{(\sqrt{\pi/2}/e)/(V_0\sqrt{k})+V_0\sqrt{k}}{1/(V_0\sqrt{k})+V_0\sqrt{k}}\right)\cdot \ln\left(1+eV_0\sqrt{k}\right). \qquad (12)$$

In contrast to the derived formula (11), this formula has a heuristic character. Here $e = 2.71828$, $k = 1/V_{term}^2$. The vertical asymptote formula from [2] is written as follows

$$x_{as} = L + \frac{1}{2gk}\ln\left[\left(e^2\frac{V_1}{V_2}\right)^{(\cos\theta_1)}(V_1\sqrt{k})^{(\cos\theta_2)}\right]. \qquad (13)$$

The following designations are introduced in formula (13):

$$L = V_a T, \quad V_a = V(0), \quad T = 2\sqrt{\frac{2H}{g}}, \quad H = \frac{V_0^2\sin^2\theta_0}{g(2+kV_0^2\sin\theta_0)}, \quad x_a = \sqrt{LH\cot\theta_0},$$

$$\theta_d = -\arctan\left[\frac{LH}{(L-x_a)^2}\right], \quad V_1 = V(\theta_d), \quad \theta_2 = (\theta_d - \pi/2)/2, \quad V_2 = V(\theta_2).$$

In this case, the entered parameters $L$, $H$, $T$, $x_a$, $\theta_d$ have the following geometric meaning (see figure 1): $H$ - the maximum height of ascent of the projectile, $x_a$ - the abscissa of the trajectory apex, $L$ - flight range, $T$ - motion time, $\theta_d$ - impact angle with respect to the horizontal.

The results of comparing the accuracy of formulas (11), (12), (13) are presented in figures 6 and 7. The movement of a baseball and a badminton shuttlecock is considered. The values of the parameters $V_0$ and $k$ are taken from [1].

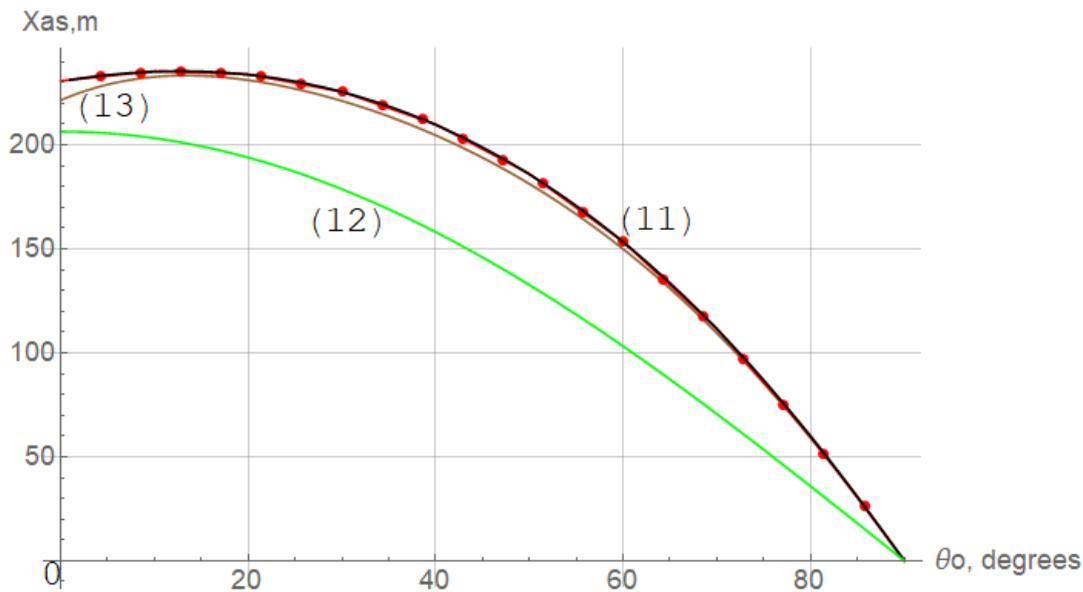

**Figure 6.** The graphs of the function $x_{as}(\theta_0)$ for baseball. The graphs are constructed for the values $V_0 = 55$ m/s, $k = 0.000625$ s$^2$/m$^2$. The value of the parameter $k$ corresponds to the value $V_{term} = 40$ m/s.

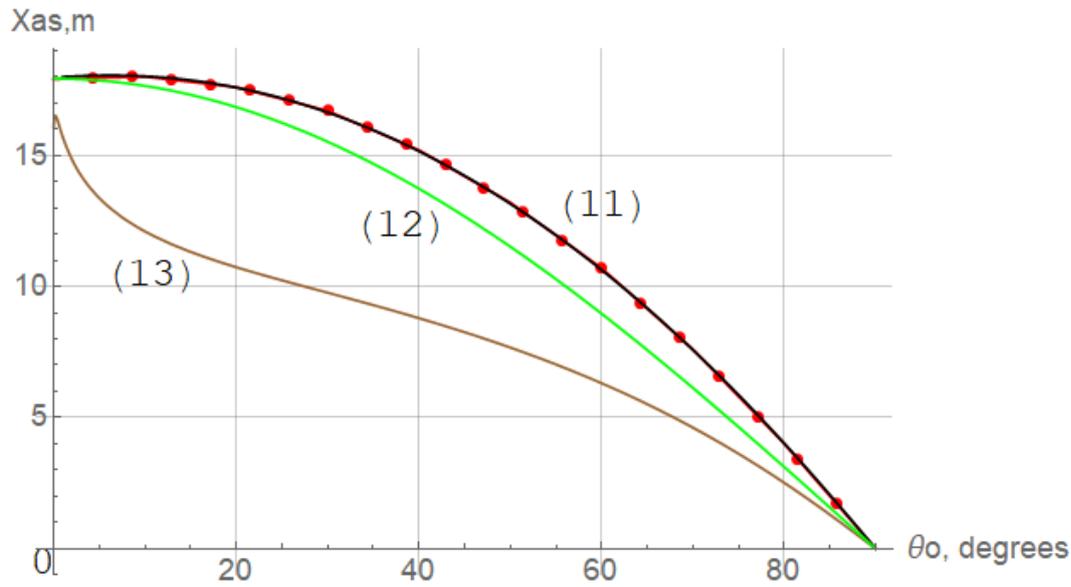

**Figure 7.** The graphs of the function $x_{as}(\theta_0)$ for badminton shuttlecock. The graphs are constructed for the values $V_0 = 117$ m/s, $k = 0.022$ s$^2$/m$^2$. The value of the parameter $k$ corresponds to the value $V_{term} = 6.7$ m/s.

Solid black curves (11) were plotted using formula (11), green curves (12) were plotted using formula (12), brown curves (13) were calculated using formula (13). The red dotted curves are obtained by integrating the system of equations (1) using the Runge-Kutta method of the 4th order. From figures 6, 7 it follows that formula (11) provides the best accuracy. Curves (11) completely coincide with the numerical solution.

Besides, we note that the obtained formulas (6)–(8) allow us to determine directly from the initial conditions $V_0$, $\theta_0$, in addition to the asymptote, other important characteristics of the projectile motion. Then

$$H = y_1(0) - y_1(\theta_0), \qquad x_a = x_1(0) - x_1(\theta_0), \qquad t_a = t_1(0) - t_1(\theta_0).$$

These values were previously found analytically in [3]. To determine the characteristics of motion $L$ and $T$, it is necessary to find impact angle with respect to the horizontal $\theta_d$ from the condition $y(\theta_d) = 0$. Then $L = x(\theta_d)$, $T = t(\theta_d)$.

Thus, the original results in this research are:

1 – analytical formulas (10) for constructing the trajectory of the projectile over the entire interval of variation of the angle of inclination ($-\pi/2 \le \theta \le \theta_0$]. The range of applicability of these formulas is not limited by any conditions (within the limits of applicability of the quadratic law of resistance of the medium). The constructed analytical solutions are universal.

2 – analytical formula (11) for the value of the vertical asymptote of the projectile trajectory. The resulting formula is superior in accuracy to all previously proposed formulas.

## 6. Conclusion

Thus, a successful approximation of the function $f(\theta)$ made it possible to calculate the integrals (3) in elementary functions and to obtain a highly accurate analytical solution of the problem of the motion of the projectile in the air over an arbitrarily large period of time. A relatively simple analytical a for the vertical asymptote of the projectile trajectory is obtained. The proposed approach based on the use of analytic formulae makes it possible to simplify significantly a qualitative analysis of the motion of a projectile with the air drag taken into account. All basic variables of the motion are described by analytical formulae containing elementary functions. Moreover, numerical values of the sought variables are determined with high accuracy. It can be implemented even on a standard calculator. Thus, proposed formulas make it possible to study projectile motion with quadratic drag force even for first-year undergraduates.